\documentstyle[12pt,aasms4]{article}   

\lefthead{Garc\'ia-Barreto et al.}
\righthead{Central Activity in the Galaxy NGC 3367}
\begin{document} 
 
\title{Central Activity in the Barred Galaxy NGC 3367}

\author{J.A. Garc\'{\i}a-Barreto\altaffilmark{1} and L. Rudnick}
\affil{Department of Astronomy, University of Minnesota, 116 Church St.,
S.E., Minneapolis, MN 55455}

\and

\author{J. Franco and M. Martos}
\affil{Instituto de Astronom\'{\i}a, Universidad Nacional Aut\'onoma de
M\'exico, Apartado Postal 70-264, M\'exico D.F. 04510 M\'exico}

\altaffiltext{1}{On a Sabbatical leave from Instituto de Astronom\'{\i}a,
Universidad Nacional Aut\'onoma de M\'exico, Apartado Postal 70-264, 
M\'exico D.F. 04510 M\'exico} 

\begin{abstract}

We report the radio continuum structure of the barred galaxy NGC 3367 with an
angular resolution of $\sim4''.5$. The radio structure indicates emission from
the disk and from a triple source consisting of the nucleus straddled by two 
extended sources (the lobes).  The 
triple source shows an excess of radio continuum emission  compared to the 
emission expected from the total radio-H$\alpha$ correlation, suggesting a 
non-thermal origin probably related to AGN activity and not to star formation 
processes. The triple source is approximately 12 kpc in extent at a P.A. 
$\approx40^{\circ}$, close (but not aligned) to that of the stellar bar, P.A. 
$\approx65^{\circ}$. Only the southwest lobe is polarized. The polarization 
asymmetry between the two lobes suggests that the triple source axis is 
slightly out of the plane. If the origin of the emission is an outflow of 
plasma 
from an AGN, similar to weak radio galaxies and NGC 1068, NGC 3367 provides an 
excellent laboratory object to study a possible interaction of the ejected 
material and the barred galaxy. 

\end{abstract}

\section{Introduction}

NGC~3367 is a face-on SBc(s) barred spiral galaxy inclined with respect to the 
plane of the sky at an angle i$\approx6^{\circ}$ (\cite{gro85}). It can be 
considered an isolated galaxy at a distance of 43.6 Mpc with a far distant 
neighbor, NGC 3419, behind the Leo Spur group of galaxies (i.e. using H$_0=75$ 
km s$^{-1}$ Mpc$^{-1}$ and assuming that the Milky Way is moving towards the 
Virgo Cluster at 300 km s$^{-1}$; \cite{tul88}). At this distance, an angular
diameter of 1$''$ corresponds to $\approx210$ pc. The stellar bar has an angular 
diameter of $\approx32''$ (6.72 kpc) oriented at P.A. $\approx65^{\circ}$, and there is a 
southwest optical structure resembling a ``bow shock'', along which lies a 
half ring of H$\alpha$ knots at a radius of about 10 kpc from the nucleus 
(\cite{gar96a}). VLA observations at 15$''$ angular resolution by 
Condon et al., (1990) at 1.46 GHz show emission from the disk as well as from 
the nucleus and from two sources straddling the nucleus. NGC~3367 is also an 
X-ray and far infrared emitter (\cite{gio90,sto91,fab92,soi89}). The X-ray
emission extends beyond the disk, but peaks some 21$''$ from the compact
nucleus in the southwest direction (Gioia et al. 1990; Stocke et al. 1991;
Fabbiano et al. 1992). This is coincident with the southwest radio continuum
lobe. Its X-ray luminosity is stronger than other normal spirals, but weaker 
than Seyfert galaxies. It also has a strong far-IR emission, of
about L$_{FIR}\approx 2\times 10^{10}$ L$_{\odot}$ (Soifer et al. 1989), with a
relatively high dust temperature of T$_D{\approx}35~$ K (Garcia-Barreto et al. 
1993). The nuclear region also shows a H$\alpha$ peak intensity of 
$2.4\times 10^{-13}$ ergs s$^{-1}$ cm$^{-2}$, with 
FWHM$\sim650$ km s$^{-1}$, and [OIII $\lambda5007$]$\approx0.5$ 
H${\beta}$ (\cite{ver86}). All these characteristics are similar to 
those found in Seyfert galaxies, and thus NGC 3367 has been classified as a 
Seyfert-like galaxy (\cite{ver86}).

Radio continuum emission from the central regions of disk galaxies has been 
detected at low resolution from almost every nearby spiral 
(\cite{hum80,gio82,con92,gar93,nik97}), and high resolution studies show that
this is usually due to circumnuclear emission, with a variety of sizes and 
morphologies. Double, triple or jet-like radio continuum structures have been 
found in several barred spirals with well-known Seyfert activity, such 
as NGC 1068, NGC 4151, and NGC 5728 but only NGC 1068, a Sy 2, displays 
extended radio continuum disk emission and has a central (0.4 kpc) source with two
lobes (\cite{wil87}), very 
much like a small-scale radio galaxy (\cite{van82,wyn85,ulv87,wil87}). 

In this paper we present new VLA\footnote{The VLA is part of the NRAO which is 
operated by Associated Universities Inc. under contract with the National 
Science Foundation.} radio continuum observations of NGC~3367 at 1.4 GHz with 
a beam of $\sim4''.5$. The radio continuum image is complex, showing emission 
from the central region, from many unresolved sources associated with star-forming regions, and from two extended regions straddling the central region. 
For the purpose of analysis in this paper we will refer to the emission from 
the central region plus the two extended sources as the {\it triple source} and 
we will refer to the emission from the disk emission as {\it star-formation} 
emission. 

\section{Observations and Results}
	
We have carried out radio continuum observations at the VLA in New Mexico 
in the B array at 1.3851 GHz and at 1.4649 GHz in 1997 May 18 using 27 antennas 
with 50 MHz bandwidths, and 7$^h$.5 integration time on NGC 3367. 
We observed 3C286 as the amplitude and polarization calibrator and 1120+143 as 
the phase calibrator. We assumed a flux density of 3C286 of S$_{\nu}=14.941$ Jy 
at ${\nu}=1.3851$ GHz, and S$_{\nu}=14.554$ Jy at ${\nu}=1.4649$ GHz. Several iterations of phase self-calibration and one iteration of amplitude  self-calibration were used. The final maps of the total intensity of NGC 3367 after 25,000 iterations using AIPS cleaning task IMAGR have an rms noise of 
$\sim18~{\mu}$Jy/beam and a cleaned beam of $4''.51 \times 4''.34$ at P.A . 
$\sim-71^{\circ}$ using uniform weighting. The final maps, however, still have  weak ripples in the east-west direction, most likely due to phase residuals.  The maps in Q and U have an rms of $\sim10~{\mu}$Jy/beam.

\subsection{Total intensity}

Figure 1 shows the total intensity map of NGC 3367. There are four different
types of regions that are apparent in the figure: a) the unresolved source from
the nuclear region; b) the northeast and the southwest extended emissions, i.e. 
the lobes, (see Fig. 2) that are connected with the nuclear region; c) many unresolved sources in the disk of the galaxy; and d) diffuse disk emission. We designate regions a) and b) as the ``triple source'' for later discussion. The total integrated flux density is 89.3 mJy, slightly less than low 
resolution (VLA-D FWHM$\approx45''$) value of 119.5 mJy, which is more sensitive to diffuse emission (\cite{con98}).  The central source is slightly resolved, with an approximate deconvolved FWHM size of $1''.7 \times 1''.6$, 
at P.A.$\approx100^{\circ}$. The integrated flux densities of the northeast and 
southwest lobes are $\approx13$ mJy and $\approx23$ mJy respectively, and 
span 12 kpc. The disk emission is generated from at least 55 unresolved sources (each with a peak flux of ${\leq}500~{\mu}$Jy/beam) and there is diffuse emission associated with each of them.

The nucleus is connected to the lobes by extensions at an angle of 
 $\approx70^{\circ}$, coincidental
with the P.A. of the innermost contours of the convolved red-continuum image 
of the stellar bar (\cite{gar96b}). A line joining the two lobes is at P.A. $\approx45^{\circ}$, 
and a one dimensional plot along this line (Figure 3) shows the brightening 
towards the outer parts of the lobe, similar to the observed behavior of
powerful radio galaxies (\cite{fer97}).  

\subsection{Polarized intensity}

Polarized emission was detected only from the southwest lobe (Figure 4). 
The fractional polarization varies through
the southwest lobe; from $\approx$4\% to $\approx$24\%. The polarization angle also varies from P.A. 
$\approx70^{\circ}$ to $\sim145^{\circ}$, with no corrections made for 
Faraday rotation. On average, the polarization in the 
northeast lobe is lower than $\approx3\%$. 

\section{The central Triple Source}

If the extended radio emission in NGC 3367 is a combination of emission related 
to star formation and to AGN activity, then, in principle, it should be 
possible to separate their contributions by using the distribution of 
H$\alpha$ emission. 

The H$\alpha$ emission and the thermal radio emission are proportional to the 
number of ionizing photons. The ionizing photons come primarily from the most
massive OB stars in stellar clusters. In addition, non-thermal emission from 
star formation regions comes from the SNR remnants. We expect a correlation 
between H$\alpha$ and radio emission, although this can be complicated by 
differences in extinctions, 
in the SNR populations, or in magnetic field strengths (e.g. 
\cite{dee97}).

Here we have adopted a more empirical approach, and have simply subtracted 
scaled versions of the H$\alpha$ map, Figure 5 (from \cite{gar96b}) from the radio map. The exact 
scaling factor is not relevant for the rest of the discussion. With the aid of 
these subtracted images (see Figure 6), we are able to eliminate the 
contributions to the radio from the brightest HII regions. Then we can get rough estimates 
of the strength and morphology of the remaining non-thermal emission which 
we identify as AGN-related.

The residual emission is dominated by the triple source previously discussed, 
along with long ridges of emission on both sides of this source (see Figure 
7). These ridges 
are likely to be disk emission, and possibly  associated with cosmic ray
acceleration in large scale shocks. They are coincident with optical low 
surface brightness regions, are relatively weak in H$\alpha$, and lie outside 
of the stellar bar. Their locations seem to correspond with locations at which
large scale shocks can be produced by gas flows driven by a bar potential (see
Roberts et al. 1979, and figure 6.28 of Binney \& Tremaine 1987). This issue
is currently under investigation with a magnetohydrodynamical code, and it will 
be addressed in a forthcoming paper. At the moment, we center the 
discussion on the AGN related emission. 

The power of the triple source is about 10$^{22}$ W Hz$^{-1}$, approximately 
half of the total radio power from NGC 3367 and similar to those of Seyfert 2 
galaxies. It is known that both Seyfert and Markarian galaxies have double or 
triple radio continuum sources (i. e., NGC 1068, Mrk3, Mrk 6, Mrk 78, NGC 4151, 
NGC 5548, and NGC 5728; \cite{ulv81,van82,ulv84,bau93}), and some edge-on 
Seyfert galaxies and starburst galaxies show large scale radio continuum 
structures along their minor axes (\cite{col96a,col96b}). In the case of barred 
spirals, the relative orientation of the bar and the double and triple sources 
varies from galaxy to galaxy. A comparison with other types of galaxies is 
presented in Section 5. 

The fact that only one lobe is polarized may indicate that they do not lie 
on the plane of the disk, but may be inclined with respect to the midplane. If 
this is the case, the emission from the far lobe (the northeast) will be 
depolarized by the disk material and only the near 
lobe (the southwest) will show some polarization. Their sizes indicate that, 
in projection, they cover about one third of the galaxy's diameter.

\section{Is NGC 3367 a transition galaxy: normal barred $\Rightarrow$
Seyfert galaxy?}
 
The optical properties of the nucleus and the extended disk of NGC 3367 are 
similar to those found in both normal spirals and weak Seyfert 2 galaxies. In 
particular, [OIII]$\approx0.5$H$\beta$, and [OI]$\approx0.3$[OIII] 
(\cite{ver86}). The nuclear optical spectrum suggest a composite spectrum between 
a normal HII spiral and a weak Liner or a weak Seyfert 2 galaxy (\cite{ver97}). 
Even though NGC 3367 does not show the optical nuclear characteristics of 
Seyfert galaxies (with broad Balmer lines with FWHM $\geq$1000 km s$^{-1}$ and 
[OIII]/H$\beta\geq$3), its radio power at 1.46GHz is stronger than Seyfert 1 galaxies and is comparable to other Seyfert 2 galaxies. 
As stated above, the power of the central source plus the lobes is P$_{1.4GHz} 
\sim 10^{22}$ W Hz$^{-1}$. Also, the global spectral index of NGC 3367 is 
$\alpha\approx-1$, from 1.4 GHz up to 4.8 GHz, similar to the spectral index found in other Seyfert galaxies (\cite{deb78,con90}) and slighty steeper than in normal spirals (\cite{kle81,con90,nik97}). 

There are several barred galaxies showing similar radio continuum morphologies, 
and NGC 1068 is probably the best known example. At a distance of 15.6 Mpc, 
the power of the inner 13$''$ structure in NGC 1068 at 6 cm is P$_{4.8GHz}\sim 
3.4\times10^{21}$ W Hz$^{-1}$ (\cite{wil87}). It also has a central source 
plus two lobes (parallel to the stellar bar), a bright stellar bar and a nearly face on disk (\cite{wil87,bal87,sco88}). Elongated radio continuum structures are relatively common in the central parts of barred galaxies with Seyfert-like activity, but the relative orientation between the structures and the corresponding bars do not show any particular type of alignment (\cite{sch88,ulv81,wil93,mun95}). A summary of the global properties of NGC 3367 and other (normal, Seyfert 2's, Seyfert 1's, and radio) galaxies is shown in Table 1.

The angular distances on the plane of the sky of NGC 3367 from the central 
source to the brightest sources in the northeast and southwest lobes are 
N-NE$\sim33''$ and N-SW$\sim26''$ respectively. These angular distances would correspond to
linear distances of $\sim7$ kpc and $\sim6$ kpc. For 
comparison, the northeast lobe in NGC 1068 is only $\sim 400$ pc from its 
nucleus. At the other extreme the distance from the center to the 
north lobe in the spiral radio galaxy 0313-192 (in the Abell 428 cluster) is  $\approx100$ kpc (\cite{led97}). The extents of the so called extranuclear 
emission observed in edge-on Seyfert galaxies range from 0.6 kpc, in NGC 4051, 
to 30 Kpc, in Mrk 231, with an average of $\sim4.5$ kpc (\cite{bau93}). Thus,
the lobes of NGC 3367 are within the wide range of sizes observed in other active
galaxies. The spatial resolution of our observations, however, is not suitable 
to analyze the structure of the connection between the lobes and the central 
nuclear source (i. e., jets, plasmoids from a galactic wind, young 
superbubbles, etc), or to analyze whether any emission originates from the 
inner regions of the bar. Also, if there is a radio continuum 
circumnuclear structure in NGC 3367, it must be less than 4$''$ in 
diameter, and higher angular resolution observations would be needed to resolve 
it. The central ionized gas structure in H$\alpha$ is also dominated by an 
unresolved bright source and weak emission extending to about a radius of 
$2''.5$ (\cite{gar96a}).

\section{The Triple Source / galaxy relation}

Several models have been proposed for the radio emission from a Seyfert nucleus
(\cite{bla79,ped85,wil87,tay89}). In our case the key questions are the origin 
of the triple radio source, its structure, and the polarization asymmetry.
 
The total gas content in NGC 3367, M$_g\sim7\times10^{9}$ M$_{\odot}$ 
(\cite{gio83,huc85}), is 
similar to the amount observed in other Sc galaxies. The AGN activity that 
originates the triple radio source in NGC 3367 was probably triggered by an
episode of efficient gas transfer to the innermost central regions. As stated
above, gas from large radii can be transfered to the central regions by the 
action of the non-axisymmetric gravitational potential (\cite{bin87,fri97}), 
or by the interaction with 
another galaxy (\cite{lyn76,too78}). In NGC 3367 both scenarios are 
possible: it has a well formed bar, and it has a string of HII regions 
(forming a semicircle) that could have originated in an off-center 
collision with an intruder (\cite{gar96a}). The bar, however, can only drive 
the gas inflow to the ILR and an additional pertubation (say, the
interaction with an intruder) is needed in order to send the gas even closer 
to the dynamical center (\cite{too78}).

On larger scales, the gas also strongly influences the appearance and structure of the triple radio source.  The polarization asymmetry is likely caused by depolarization through an ionized component, such as that seen in X-rays.  The limb-brightened outer edges indicate an interaction with gaseous
external medium.  Similarly, the S-shape of the triple source could result from shears in galactic rotation or gradients in the ambient medium (e.g., \cite{hen81}).  Precession of the jet due to torques from infalling material (\cite{toh82}) could also play a role.

The interactions between the radio plasma and the other gaseous components are worthy of further study, such as searching for molecular and shocked gas at
the leading edges or where the jets from the nucleus make large changes in direction.  Numerical simulations of jet flows in a barred galaxy would also be of interest.

\section {Conclusions}

We have observed, with the VLA B array, the radio continuum emission from the 
barred galaxy NGC~3367 with 4$''.5$ spatial resolution. The radio continuum 
emission maps shows emission from four distinct regions: a) star forming
regions, b) the nucleus, and c) two extended,
symmetric, lobes connected with the nuclear source, and d) diffuse emission 
from the disk that is not coincident with bright H$\alpha$ regions. Our main 
results are:
\begin{itemize}
\item The optical and radio characteristics suggest that NGC 3367 is a 
transition galaxy between  normal to Seyfert.
\item The total radio continuum emission from regions of star formation 
correlates with the observed H$\alpha$ emission, in agreement with a similar 
correlation found in other galaxies.
\item The strongest radio continuum emission is from a barely resolved nuclear
source and two extended regions symmetrically located opposite to the nucleus, 
forming what we call a triple source with total extent $\sim 12$ kpc. The
morphology of the triple source is similar to those observed in radio galaxies,
and to the structure seen in the inner 0.4 kpc in NGC 1068. 
\item The triple source could be the result of nuclear ejecta in the northeast 
and southwest directions, and inclined with respect to the galactic plane. 
The $S$-shape of the lobes could be due to interaction with the ambient gas.
\item Only the southwest lobe is found to be polarized. The northeast lobe is 
probably de-polarized as the beam passes through a randomly oriented magnetic 
field in the disk of the galaxy.

\end{itemize}

\section*{Acknowledgements}
We would like to thank the anonymous referee for useful comments and 
suggestions on how to improve the paper. It is a pleasure to thank Dr. Barry Clark and Dr. W. Miller Goss for the VLA 
allocated observing time. We would like to thank C. Lacey for useful comments. JAG-B wishes to thank the assistance and help of 
Greg Taylor for observing file preparation and calibration of the VLA data. 
The work of JF and MM was partially supported by a DGAPA-UNAM grant, CONACyT grants 400354-5-4843E and 400354-5-0639PE, and by a R\&D CRAY Research grant.
JAG-B acknowledges partial financial support from DGAPA-UNAM and by CONACYT 
(Mexico) that enable him to work during his sabbatical year at the Department 
of Astronomy of the University of Minnesota where most of the analysis was 
done. L.R. extragalactic research at Minnesota is supported by the National 
Science Foundation through grant NSF-AST 9616984.

\normalsize

\clearpage

\figcaption{Radio continuum emission from NGC 3367 with a beam FWHM$\approx 
4''.5\times4''.3$ at PA$\approx-70^{\circ}$. The contours are in units of 
18$\mu$Jy/beam and the levels are -3, 3, 5, 7, 9, 11, 13, 15, 17, 20, 25, 35, 50, 
65, 80, 110, 120, 140 and 160. The weak extension from the nuclear source into 
the southeast and the weak extensions to the west of the galaxy are probably 
not real but related to phase problems in the data.  \label{fig1}}

\figcaption{Radio continuum emission from the extended region northeast and 
southwest of the nucleus. The contours are in units of 18$\mu$Jy/beam and the levels 
are -3, 7, 11, 15, 19, 21.5, 23, 25, 27, 30, 40, 55, 70, 90, 120, 140 and 160. 
 \label{fig2}}

\figcaption{One dimensional plot of the total radio continuum emission along 
the line at PA$\sim45^{\circ}$ (with center at $\alpha=10^h46^m35^s.1$, 
$\delta=13^{\circ}45'04''.8$. Notice that the flux density increases with
distance from the nucleus, and has a sharp drop at the end of the lobes. The 
large central flux is from the nuclear region. Negative distances are 
towards southwest lobe. \label{fig3}}

\figcaption{Electric field vectors plotted over the total intensity map of the 
southwest lobe in NGC 3367. The contours are similar to the contours in Fig. 1. 
 The length of the 
vector is proportional to the polarized brightness with 1$''$ indicating 
200$\mu$Jy/beam. \label{fig4}}

\figcaption{a) H$\alpha$+[NII] continuum-free image of NGC 3367 
(Garcia-Barreto, Franco \& Carrillo 1996; Garcia-Barreto et al., 1996) 
convolved to a gaussian beam as the clean beam in the radio continuum 
map (see Fig. 1). The contour units are $6\times10^{-16}$ ergs s$^{-1}$ 
cm$^{-2}$ arcsec$^{-2}$ and the levels are -3, 3, 5, 7, 9, 11, 13, 15, 17, 
20, 25, 30, 35, 50, 65, 80, 100, 120, 140 and 160. Notice the limb brightened 
edge forming a semicircle in the southwestern part of the galaxy. b) Radio 
continuum emission contours superimposed on a grey scale image of the 
H$\alpha$+[NII] continuum-free emission from NGC 3367. 
The contour units and levels are similar to the contours in Figure 1.  \label{fig5}}

\figcaption{Map of the radio continuum emission minus a scaled H$\alpha$ image. 
The scale was $7.2\times10^9$. Notice : i) more emission from weak HII region 
dissapeared at the northwest, ii) the streamer starting at the norteast is 
getting narrower than the emission from the original radio map (see 
Fig. 1) and iii) a limb brightened eastern edge of the northeast lobe 
is more pronounced than the emission from the original radio map (see 
Fig. 1). In this map there is negative emission from the bright 
H$\alpha$ regions in the south and west of the galaxy indicating that the 
subtraction was overdone in those areas. The radio continuum emission seen in this map is most likely 
not directly related to star-formation. 
\label{fig6}}

\figcaption{Optical broadband I ($\lambda\sim8040$ \AA) image 
(from Garcia-Barreto et al. 1996b) in contours 
superimposed on the total radio continuum emission. Contours are such as to 
mainly indicate the extent of the stellar bar and the spatial extent of the 
disk. The grey scale was chosen such as to show the emission mainly from the 
lobes. Notice that the innermost radio continuum emission originates in the 
same position angle as the stellar bar and then slowly the emission 
deviates from this PA to smaller values in the northeast. \label{fig7}} 

\clearpage


\begin{deluxetable}{llllllll}
\footnotesize
\tablecaption{Comparison Properties of NGC 3367 and other Galaxies. 
\label{tbl-4}}
\tablewidth{0pt}
\tablehead{
   & \colhead{Normal}   & \colhead{NGC 3367} & \colhead{NGC 1068} 
& \colhead{Seyfert 2's} & \colhead{Seyfert 1's} & \colhead{Radio G} 
& \colhead{Refs.}
}
\startdata
\\ 
$\lambda~5007$/H$\beta$  & $\leq~1$  & 0.4 & $\geq~2$ & $\geq~1$ & $\geq~1$ 
& $\geq~1$ & 1, 2, 17\tablenotemark{a}
\nl
L$_{[OIII]}$ W & $\sim10^{28}$ & $4~10^{32}$ & $\sim10^{33}$ & 10$^{34}-10^{37}$ & 10$^{34}-10^{37}$ & 10$^{34}$ & 1,2,3,17
\nl
q(FIR/Radio)     & $\sim2.3$   & 2 - 2.4     & 1.7 & 1.5 - 2.3 & 2 - 2.4 & $\leq~1$  & 4,5,6,7 \nl
Log P$_{1.4GHz}$ & $\leq20$  & $22.3$ & $23$ & $21 - 23.5$ & $20 - 22$ & $\geq~24$ & 3,6,7,8
\nl
Dimensions\tablenotemark{b} & $\leq1$ kpc & $5.5 - 7$ kpc & $\sim0.45$ kpc 
& $0.5 - 5$ kpc & $\leq100$ pc & $\geq100$ kpc & 6,7,8,9,10,11,12
\nl
log L$_X$ ergs s$^{-1}$ & $\sim38.5$ & 40.9 & 41.74 & $\leq~42$ & $\leq~43$ & 40 - 44 
& 13,14,15,16
\nl
\enddata
\tablenotetext{a}{References: 1)\cite{bal87}, 2)\cite{ver86}, 3)\cite{wil96},
4)\cite{con91}, 5)\cite{led97}, 6) this paper, 7)\cite{con90}, 8)\cite{ulv84}, 
9)\cite{hum87}, 10)\cite{wil87}, 11)\cite{bau93}, 12)\cite{col96b}, 13)\cite{fab92}, 14)\cite{dul96}, 15)\cite{wil91}, 16)\cite{rhe94}, 
17)\cite{law96}} 
\tablenotetext{b}{Distances from nucleus to circumnuclear region in normal galaxies, from nucleus to lobe in NGC 1068, 3367, Seyferts and Radio galaxies}
\end{deluxetable}

\clearpage

\clearpage

\end{document}